\documentclass[a4paper,english]{lipics-v2019}
\usepackage{siunitx}
\nolinenumbers

\newcommand{\BWT}{\ensuremath{\mathrm{BWT}}}
\newcommand{\LCE}{\ensuremath{\mathrm{LCE}}}
\newcommand{\LCP}{\ensuremath{\mathrm{LCP}}}
\newcommand{\PFP}{\ensuremath{\mathrm{PFP}}}
\newcommand{\SA}{\ensuremath{\mathrm{SA}}}
\newcommand{\rank}{\ensuremath{\mathrm{rank}}}
\newcommand{\select}{\ensuremath{\mathrm{select}}}

\title{PFP Data Structures}

\author{Christina Boucher}{University of Florida, USA}{}{}{}

\author{Ond\v{r}ej Cvacho}{Czech Technical University, Czech Republic}{}{}{}

\author{Travis Gagie}{Dalhousie University, Canada}{}{}{}

\author{Jan Holub}{Czech Technical University, Czech Republic}{}{}{}

\author{Giovanni Manzini}{University of Eastern Piedmont, Italy}{}{}{}

\author{Gonzalo Navarro}{University of Chile}{}{}{}

\author{Massimiliano Rossi}{University of Florida, USA}{}{}{}

\authorrunning{C. Boucher, O. Cvacho, T. Gagie, J. Holub, G. Manzini, G. Navarro and M. Rossi}

\Copyright{Christina Boucher, Ond\v{r}ej Cvacho, Travis Gagie, Jan Holub, Giovanni Manzini, Gonzalo Navarro and Massimiliano Rossi}



\acknowledgements{The authors thank Manuel C{\'a}ceres for helping with the code of the block-tree compressed suffix tree.} 

\EventEditors{John Q. Open and Joan R. Access}
\EventNoEds{2}
\EventLongTitle{42nd Conference on Very Important Topics (CVIT 2016)}
\EventShortTitle{CVIT 2016}
\EventAcronym{CVIT}
\EventYear{2016}
\EventDate{December 24--27, 2016}
\EventLocation{Little Whinging, United Kingdom}
\EventLogo{}
\SeriesVolume{42}
\ArticleNo{23}

\begin{document}

\maketitle
\begin{abstract}
Prefix-free parsing (PFP) was introduced by Boucher et al. (2019) as a preprocessing step to ease the computation of Burrows-Wheeler Transforms (BWTs) of genomic databases.  Given a string $S$, it produces a dictionary $D$ and a parse $P$ of overlapping phrases such that $\BWT (S)$ can be computed from $D$ and $P$ in time and workspace bounded in terms of their combined size $|\PFP (S)|$.  In practice $D$ and $P$ are significantly smaller than $S$ and computing $\BWT (S)$ from them is more efficient than computing it from $S$ directly, at least when $S$ consists of genomes from individuals of the same species.  In this paper, we consider $\PFP (S)$ as a {\em data structure} and show how it can be augmented to support the following queries quickly, still in $O (|\PFP (S)|)$ space: longest common extension (LCE), suffix array (SA), longest common prefix (LCP) and BWT.  Lastly, we provide experimental evidence that the PFP data structure can be efficiently constructed for very large repetitive datasets: it takes one hour and 54 GB peak memory for $1000$ variants of human chromosome 19, initially occupying roughly 56 GB.
\end{abstract}

\section{Introduction}
\label{sec:old_introduction}

There have been dozens of algorithms proposed for building suffix arrays (SAs), Burrows-Wheeler Transforms (BWTs), longest common prefix (LCP) arrays, FM-indexes and similar data structures, but not many of them are well-suited to dealing with genomic databases.  A thousand human genomes, for example, occupy a few gigabytes when compressed properly but a few terabytes uncompressed, so construction algorithms that cannot be parallelized or that use many random accesses to the raw data do not run well on commodity computers.  A hundred thousand human genomes take hundreds of terabytes uncompressed, so then such algorithms can run on only handful of computers in the world, if at all.  Construction has been an important issue in indexing for nearly fifty years at least, since Weiner's~\cite{DBLP:conf/focs/Weiner73} breakthrough showing that suffix trees can be built in linear time and space.  As Ferragina et al.~\cite{DBLP:journals/algorithmica/FerraginaGM12} pointed out, ``to use [an index] one must first {\em build} it!''  Computer memories are getting larger every day, of course, but not as quickly as genomic databases: the 1000 Genomes Project Consortium announced the sequencing of 1092 human genomes in 2012 and Genomics England announced the sequencing of 100K human genomes in 2018, significantly outpacing Moore's Law.

Suppose we want to index a thousand human genomes in such a way that we can support standard bioinformatics tasks such as DNA sequence read alignment (see~\cite{DBLP:journals/bib/Consortium18} for more discussion).  For example, given a sequence read, we might want to determine whether and where it occurs in the database and, if it does not occur, then which of its substrings occur and where.  Determining whether and where the sequence read occurs is a question of exact pattern matching and there has already been significant progress towards solving that problem, both in theory~\cite{gagie2020fully} and in practice~\cite{DBLP:journals/almob/BoucherGKLMM19,kuhnle2020efficient,mun2020matching}.  Determining which of the read's substrings occur and where is apparently more difficult but potentially more interesting, because it is an important step in the seed-and-extend approach to sequence read alignment.

In order to address these issues of constructing an index that scales to thousands of genomes, Gagie et al.~\cite{gagie2020fully} designed a run-length compressed FM-index, called the $r$-index, that takes space proportional to the number $r$ of runs in the BWT and still supports fast locating queries.  In a series of papers, including Boucher et al. \cite{DBLP:journals/almob/BoucherGKLMM19} and Kuhnle et al. \cite{kuhnle2020efficient},  it was then revealed how to build instances of the $r$-index for non-trivial genomic databases.  To do this, the authors introduced a preprocessing step called {\em prefix-free parsing} (PFP), that roughly compresses the data in such a way that we can compute the BWT and SA sample directly from the compressed representation. Although this work is one step towards read alignment, it does not solve the problem entirely since Gagie et al.'s index supports only exact pattern matching and thus cannot easily be used for read alignment.

Bannai et al.~\cite{DBLP:journals/tcs/BannaiGI20} began addressing this conceptual gap by describing an efficient and relatively simple data structure which, when combined with fast random access to a genomic database, allows us compute the matching statistics of a sequence read with respect to the database. However, building Bannai et al.'s structure requires computing some new information: for each consecutive pair of runs of the same character in the BWT of the database, we need to find a minimum in the longest common prefix (LCP) interval between them.  Under normal circumstances that would not be a problem, but our options are more limited when dealing with genomic databases.  As mentioned above, we should avoid algorithms that cannot be parallelized or that use many random accesses to the raw data.  Furthermore, some constructions use several auxiliary data structures that each take space only slightly sublinear in the size of the database, and their combined space can overflow the internal memory. 

Thus, although Bannai et al.'s new structure uses space bounded in terms only of the number $r$ of runs in the BWT of the database -- with no slightly sublinear dependence on the database's raw size -- it is still unknown how to build it efficiently in practice.  Compressed suffix trees (CSTs) support queries allowing the construction of Bannai et al.'s data structure, but existing implementations, such as the CST included in the Succinct Data Structure Library (SDSL)~\cite{gbmp2014sea} and block-tree based CST (BT-CST)~\cite{DBLP:conf/spire/CaceresN19}, require too much time or memory to build when the datasets are very large.

In this paper, we take a step towards realizing the structure of Bannai et al.\ by showing how PFP can be viewed as a {\em data structure} to support the following queries quickly: longest common extension (LCE), suffix array (SA), longest common prefix (LCP) and BWT.  Although the resulting data structures are not as small as other data structures supporting the same queries, their construction time and peak memory are smaller than those of the existing data structures. This allows us to support random access SA and LCE and LCP queries for very large datasets. For example, PFP data structures can be built for $1000$ distinct variants of human chromosome 19 in slightly more than $1$ hour using $54$ GB of internal memory, that is almost the size of the raw data. With the same amount of internal memory, the other data structures cannot be built for more than $32$ distinct variants.  For the sake of brevity we assume readers are familiar with SAs, BWTs, wavelet trees, FM-indexes, etc., and their use in bioinformatics; if not, we refer them to M\"akinen et al.'s and Navarro's books~\cite{DBLP:books/cu/MBCT2015,DBLP:books/daglib/0038982}.

\section{PFP}
\label{sec:PFP}

To compute a PFP of $S [0..n - 1]$, conceptually we choose a subset of all possible strings of some length $w$, with the chosen strings called {\em trigger strings}, and then divide $S$ into overlapping phrases such that each starts with a trigger string (except possibly the first), ends with a trigger string (except possibly the last), and contains no other trigger string.

In practice we choose the trigger strings implicitly, by choosing a Karp-Rabin hash function and a parameter $p$ and passing a sliding window of length $w$ over $S$, putting a phrase break wherever the hash of the contents of the window is congruent to 0 modulo $p$ (with the contents of the window there becoming the last $w$ characters of the previous phrase and the first $w$ character of the next one).

PFP is inspired by {\tt rsync}~\cite{Tridgell1999} and {\tt spamsum} (\url{https://www.samba.org/ftp/unpacked/junkcode/spamsum/README}\,; see also~\cite{DBLP:journals/di/Kornblum06}), which have been in popular use for about twenty years.  In some cases it works badly --- e.g., if $S$ is unary then either we split it into $n - w + 1$ phrases or we do not split it at all --- but we usually end up with a parse consisting of roughly $n / p$ phrases of length roughly $p$.

It seems plausible that PFP can be adapted to have good worst-case bounds, possibly by combining it either with string synchronizing sets~\cite{DBLP:conf/stoc/KempaK19} or locally consistent parsing~\cite{DBLP:conf/soda/BirenzwigeGP20}, but this would probably make it impractical.  As it is, the parsing uses only sequential access and small workspace, so it runs well even in external memory, and it can easily be parallelized.  When $S$ consists of genomes from individuals of the same species, then the genomes are parsed roughly the same way, so the total length of the strings in the dictionary of distinct phrases can be significantly less than the total length of the genome.

In this paper we assume we have already computed for $S$ a PFP parse $P$ with dictionary $D$, using Boucher et al.'s implementation, and we now restrict ourselves to using memory proportional to their combined size $|\PFP (S)|$.  We say a phrase $S [i..j]$ {\em contains} a character $S [k]$ if $i \leq k \leq j - w$.  Notice that, since consecutive phrases overlap by $w$ characters, each character of $S$ is contained in this sense in exactly one phrase, except the last $w$ characters of $S$.  To simplify the presentation, assume $S$ is cyclic and starts with a trigger string --- if need be, we can prepend one --- so each character of $S$ is contained in exactly one phrase, with no exceptions.

For example, consider the string $ \mathtt{GATTACAT\#GATACAT\#GATTAGATA}$ containing the trigger strings $\mathtt{AC}$, $\mathtt{AG}$ and $\mathtt{T\#}$ of length $w = 2$.  We append $w = 2$ copies of $\#$ and consider the string as cyclic,
\[S = \mathtt{GATTACAT\#GATACAT\#GATTAGATA\#\#GATTACAT\#GATACAT\#GATTAGATA\#\#}\ldots\]
of length $n = 28$, and treat $\mathtt{\#\#}$ as a trigger string as well.  Therefore, the parse is
\begin{eqnarray*}
P & = & \mathtt{\#\#GATTAC}, \mathtt{ACAT\#}, \mathtt{T\#GATAC}, \mathtt{ACAT\#}, \mathtt{T\#GATTAG}, \mathtt{AGATA\#\#}\\
& = & D [0], D [1], D [3], D [1], D [4], D [2]
\end{eqnarray*}
and the dictionary is
\[D = \{\mathtt{\#\#GATTAC}, \mathtt{ACAT\#}, \mathtt{AGATA\#\#}, \mathtt{T\#GATAC}, \mathtt{T\#GATTAG}\}\,.\]
Notice the phrase $D [1] = \mathtt{ACAT\#}$ occurs twice in $P$.

The most important property of a prefix-free parse is, as one would expect, that it is prefix-free.  In particular, no proper phrase suffix of length at least $w$ is a prefix of any other proper phrase suffix of length at least $w$.  To see why, consider that each proper phrase suffix (i.e., a phrase suffix that is not a complete phrase) of length at least $w$ ends with a trigger string and contains no other complete trigger string.  Therefore, if a proper phrase suffix $\alpha$ of length at least $w$ is a prefix of another such phrase suffix $\beta$, then $\alpha = \beta$.

\begin{lemma}[\cite{DBLP:journals/almob/BoucherGKLMM19}]
\label{lem:prefix-freeness}
The distinct proper phrase suffixes of length at least $w$ are a prefix-free set of strings.
\end{lemma}

A useful corollary of this is that each character $S [i]$ immediately precedes in $S$ an occurrence of exactly one proper phrase suffix of length at least $w$, which is the suffix following $S [i]$ in the phrase containing it.

\begin{corollary}[\cite{DBLP:journals/almob/BoucherGKLMM19}]
\label{cor:partition}
We can partition $S$ into subsequences such that the characters in the $i$th subsequence precede in $S$ occurrences of the lexicographically $i$th proper phrase suffix of length at least $w$.
\end{corollary}

Boucher et al.\ used this corollary as a starting point for building the BWT of $S$: for each proper phrase suffix $\alpha$ of length at least $w$ that is preceded by only one distinct character $c$ in $D$, they found the beginning of the interval for $\alpha$ in the BWT by summing up the frequencies in $P$ of phrases ending with proper phrase suffixes of length at least $w$ lexicographically less than $\alpha$, then filled in the interval for $\alpha$ with as many copies of $c$ as there are phrases in $P$ ending with $\alpha$.

To fill in the BWT intervals for a proper phrase suffix $\beta$ of length at least $w$ preceded by more than one distinct character in $D$, Boucher et al.\ used the following lemma, which is easily proven by induction.  Essentially, they considered the phrases ending with $\beta$ in the order they appear in the BWT of $P$ (viewed as a sequence of lexicographically-sorted phrase identifiers), since the lemma means they are sorted by the suffixes that follow them in $S$.

\begin{lemma}[\cite{DBLP:journals/almob/BoucherGKLMM19}]
\label{lem:comparison}
Suppose $S [i..n - 1]$ and $S [j..n - 1]$ are suffixes of $S$ starting at the beginning of occurrences of trigger strings, and let $P_i$ and $P_j$ be the parses of those suffixes with each phrase represented by its lexicographic rank in $D$.  Then $S [i..n - 1]$ is lexicographically less than $S [j..n - 1]$ if and only if $P_i$ is lexicographically less than $P_j$.
\end{lemma}

\section{Components}
\label{sec:components}

First, we store $P$ and $D$ compactly but such that we can support fast random access to them.  We expect the first-order empirical entropy of $P$ to be low for the cases that interest us most --- e.g., if $S$ is a genomic database then knowing one phrase will usually let us guess the preceding phrase pretty well --- and there are practical data structures that take advantage of this~\cite{DBLP:journals/tcs/FerraginaV07,GIMNSST20}.

Second, we store a cyclic bitvector $B_P [0..n - 1]$ with a 1 marking the position of the first character in each trigger string in $S$.  We can find the index of the phrase containing a character $S [i]$ with a rank query, modulo the number of phrases in $P$, and then find the offset of $S [i]$ in that phrase with a select query and a subtraction.  Symmetrically, if we know the index of the phrase containing a character and its offset in that phrase, we can find the character's position in $S$.  For our example $\ldots \mathtt{\#\#GATTACAT\#GATACAT\#GATTAGATA\#\#}\ldots$ we store
\[B_P = 0000100100001001000001000010\,.\]
Notice that, because the bitvector is cyclic and it is convenient for the bits to align with the corresponding characters, the 1 marking the first character of the trigger string at the beginning of the first phrase is the penultimate bit.

Third, we store a bitvector $B_{\BWT} [0..n - 1]$ that, for each distinct proper phrase suffix of length at least $w$, has a 1 marking the position of the first character in the BWT of $S$ that immediately precedes in $S$ an occurrence of that phrase suffix.  Recall that, by Corollary~\ref{cor:partition}, every character of $S$ precedes an occurrence of exactly one such phrase suffix.  Table~\ref{tab:BBWT} shows that for our example
    \[B_{\BWT} = 1111110110111110111110110111\,.\]

\begin{table}[p]
\resizebox{\textwidth}{!}
{\rotatebox{-90}
{\tt \begin{tabular}{ccl}
$B_{\BWT} [i]$ & $\BWT [i]$ & $S [\SA [i]..n - 1]$\\
\hline
\rm 1 & A  & \textcolor{red}{\#\#}\\
\rm 1 & T  & \textcolor{red}{\#GATAC}AT\#GATTAGATA\#\#\\
\rm 1 & \# & \textcolor{red}{\#GATTAC}AT\#GATACAT\#GATTAGATA\#\#\\
\rm 1 & T  & \textcolor{red}{\#GATTAG}ATA\#\#\\
\rm 1 & T  & \textcolor{red}{A\#\#}\\
\rm 1 & T  & \textcolor{red}{AC}AT\#GATACAT\#GATTAGATA\#\#\\
\rm 0 & T  & \textcolor{red}{AC}AT\#GATTAGATA\#\#\\
\rm 1 & T  & \textcolor{red}{AG}ATA\#\#\\
\rm 1 & C  & \textcolor{red}{AT\#}GATACAT\#GATTAGATA\#\#\\
\rm 0 & C  & \textcolor{red}{AT\#}GATTAGATA\#\#\\
\rm 1 & G  & \textcolor{red}{ATA\#\#}\\
\rm 1 & G  & \textcolor{red}{ATAC}AT\#GATTAGATA\#\#\\
\rm 1 & G  & \textcolor{red}{ATTAC}AT\#GATACAT\#GATTAGATA\#\#\\
\rm 1 & G  & \textcolor{red}{ATTAG}ATA\#\#\\
\rm 1 & A  & \textcolor{red}{CAT\#}GATACAT\#GATTAGATA\#\#\\
\rm 0 & A  & \textcolor{red}{CAT\#}GATTAGATA\#\#\\
\rm 1 & A  & \textcolor{red}{GATA\#\#}\\
\rm 1 & \# & \textcolor{red}{GATAC}AT\#GATTAGATA\#\#\\
\rm 1 & \# & \textcolor{red}{GATTAC}AT\#GATACAT\#GATTAGATA\#\#\\
\rm 1 & \# & \textcolor{red}{GATTAG}ATA\#\#\\
\rm 1 & A  & \textcolor{red}{T\#}GATACAT\#GATTAGATA\#\#\\
\rm 0 & A  & \textcolor{red}{T\#}GATTAGATA\#\#\\
\rm 1 & A  & \textcolor{red}{TA\#\#}\\
\rm 1 & T  & \textcolor{red}{TAC}AT\#GATACAT\#GATTAGATA\#\#\\
\rm 0 & A  & \textcolor{red}{TAC}AT\#GATTAGATA\#\#\\
\rm 1 & T  & \textcolor{red}{TAG}ATA\#\#\\
\rm 1 & A  & \textcolor{red}{TTAC}AT\#GATACAT\#GATTAGATA\#\#\\
\rm 1 & A  & \textcolor{red}{TTAG}ATA\#\#
\end{tabular}}}
\medskip
\caption{$B_{\BWT}$ for our example, with the BWT of $S$ and the suffixes of $S$ in lexicographic order.  We have highlighted in red the unique proper phrase suffix of length at least $w$ following each character, to clarify how $B_{\BWT}$ is defined.  (We show $S [n - 1] = \mathtt{\#}$ and the empty suffix as {\tt \textcolor{red}{\#GATTAC}AT\#GATACAT\#GATTAGATA\#\#} and {\tt \textcolor{red}{GATTAC}AT\#GATACAT\#GATTAGATA\#\#} instead, because we consider $S$ to be cyclic and this should make clearer how the characters in the BWT are sorted.)}
\label{tab:BBWT}
\end{table}

We do not need to build the BWT of $S$ in order to build $B_{\BWT}$.  Instead, we append a unique terminator symbol to each phrase in $D$; build the suffix array and LCP array for $D$ with those terminators; tag each suffix with the frequency in $P$ of the phrase containing that suffix; and then scan the arrays, ignoring the suffixes that are whole phrases or shorter than $w$ (ignoring the terminators) and aggregating the frequencies of the suffixes that differ only by their terminators.  Table~\ref{tab:buildingBBWT} shows how we build $B_{\BWT}$ for our example.

\begin{table}[p]
\centering
\begin{tabular}{rlc@{\hspace{8ex}}rlc@{\hspace{8ex}}rlc@{\hspace{8ex}}clc}
 1 & \tt \#\#2 & 1 & 		 - & \tt ACAT\#1 & 2 &		10 & \tt CAT\#1 & 2 &		 1 & \tt TAC0 & 1\\
 - & \tt \#\#GATTAC0 & 1 &	 1 & \tt AG4 & 1 &			 - & \tt G4 & 1 &			 0 & \tt TAC3 & 1\\
 - & \tt \#1 & 2 &			 - & \tt AGATA\#\#2 & 1 &	 1 & \tt GATA\#\#2 & 1 &	 1 & \tt TAG4 & 1\\
 - & \tt \#2 & 1 &			10 & \tt AT\#1 & 2 &		 1 & \tt GATAC3 & 1 &		 1 & \tt TTAC0 & 1\\
 1 & \tt \#GATAC3 & 1 &		 1 & \tt ATA\#\#2 & 1 &		 1 & \tt GATTAC0 & 1 &		 1 & \tt TTAG4 & 1\\
 1 & \tt \#GATTAC0 & 1 &	 1 & \tt ATAC3 & 1 &		 1 & \tt GATTAG4 & 1 &		 &&\\
 1 & \tt \#GATTAG4 & 1 &	 1 & \tt ATTAC0 & 1 &		10 & \tt T\#1 & 2 &			 &&\\
 1 & \tt A\#\#2 & 1 &		 1 & \tt ATTAG4 & 1 &		 - & \tt T\#GATAC3 & 1 &	 &&\\
 1 & \tt AC0 & 1 &			 - & \tt C0 & 1 &			 - & \tt T\#GATTAG4 & 1 &	 &&\\
 0 & \tt AC3 & 1 &			 - & \tt C3 & 1 &			 1 & \tt TA\#\#2 & 1 &		 &&
\end{tabular}
\medskip
\caption{Suppose we append a unique terminator symbol to each phrase in $D$; sort the phrase suffixes {\bf (center column)}; tag each suffix with the frequency in $P$ of the phrase containing that suffix {\bf (right column)}; mark with copies of - the suffixes which are whole phrases or shorter than $w$ (ignoring the terminators), with 1 the first copy of each suffix (ignoring terminators) and with 0s the other copies {\bf (left column)}; and then append to each 1 and 0 as many copies of 0 as the phrase frequency, minus 1  {\bf (right column)}.  Then the concatenation of the 0s and 1s is $B_{\BWT}$, which is 1111110110111110111110110111 in this example.}
\label{tab:buildingBBWT}
\end{table}

Fourth, we build a table $M$ such that $M [i]$ tells us 
\begin{enumerate}
\item the length of the lexicographically $i$th proper phrase suffix $\alpha$ of length at least $w$,
\item the lexicographic range of the reversed phrases starting with $\alpha$ reversed.
\end{enumerate}
We compute the lengths while building $B_{\BWT}$, and the lexicographic range by reversing the phrases and sorting them.  Table~\ref{tab:M} shows $M$ for our example, for which the reversed phrases are $\mathtt{\#\#ATAGA}$, $\mathtt{\#TACA}$, $\mathtt{CATAG\#T}$, $\mathtt{CATTAG\#\#}$ and $\mathtt{GATTAG\#T}$.  The lexicographic range  of {\tt CA} in Table~\ref{tab:M} is $[2,3]$ since the two reversed phrases starting with CA are in positions 2 and 3 in this sorted list (counting from 0).

\begin{table}[tb]
\centering
\begin{tabular}{lcl@{\hspace{7ex}}lcl@{\hspace{7ex}}lcl@{\hspace{7ex}}lcl}
$\mathtt{\#\#}$     & 2 & $[0]$    & $\mathtt{GA}$      & 2 & $[4]$ &
$\mathtt{\#TAC}$    & 4 & $[1]$    & $\mathtt{\#\#AT}$  & 4 & $[0]$\\
$\mathtt{CATAG\#}$  & 6 & $[2]$    & $\mathtt{\#TA}$    & 3 & $[1]$ &
$\mathtt{\#\#ATAG}$ & 6 & $[0]$    & $\mathtt{CAT}$     & 3 & $[2, 3]$\\
$\mathtt{CATTAG\#}$ & 7 & $[3]$    & $\mathtt{\#\#ATA}$ & 5 & $[0]$ &
$\mathtt{CATAG}$    & 5 & $[2]$    & $\mathtt{GAT}$     & 3 & $[4]$\\
$\mathtt{GATTAG\#}$ & 7 & $[4]$    & $\mathtt{CATA}$    & 4 & $[2]$ &
$\mathtt{CATTAG}$   & 6 & $[3]$    & $\mathtt{CATT}$    & 4 & $[3]$\\
$\mathtt{\#\#A}$    & 3 & $[0]$    & $\mathtt{CATTA}$   & 5 & $[3]$ &
$\mathtt{GATTAG}$   & 6 & $[4]$    & $\mathtt{GATT}$    & 4 & $[4]$\\
$\mathtt{CA}$       & 2 & $[2, 3]$ & $\mathtt{GATTA}$   & 5 & $[4]$ &
$\mathtt{\#T}$      & 2 & $[1]$    &                    &
\end{tabular}
\medskip
\caption{The reversed proper phrase suffixes of length at least $w$ {\bf (left column)}, their lengths {\bf (center column)}, and the lexicographic range of the reversed suffixes starting with those reversed proper phrase suffixes {\bf (right column)}.}
\label{tab:M}
\end{table}

\begin{figure}[t]
\centering
\includegraphics[width=.7\textwidth]{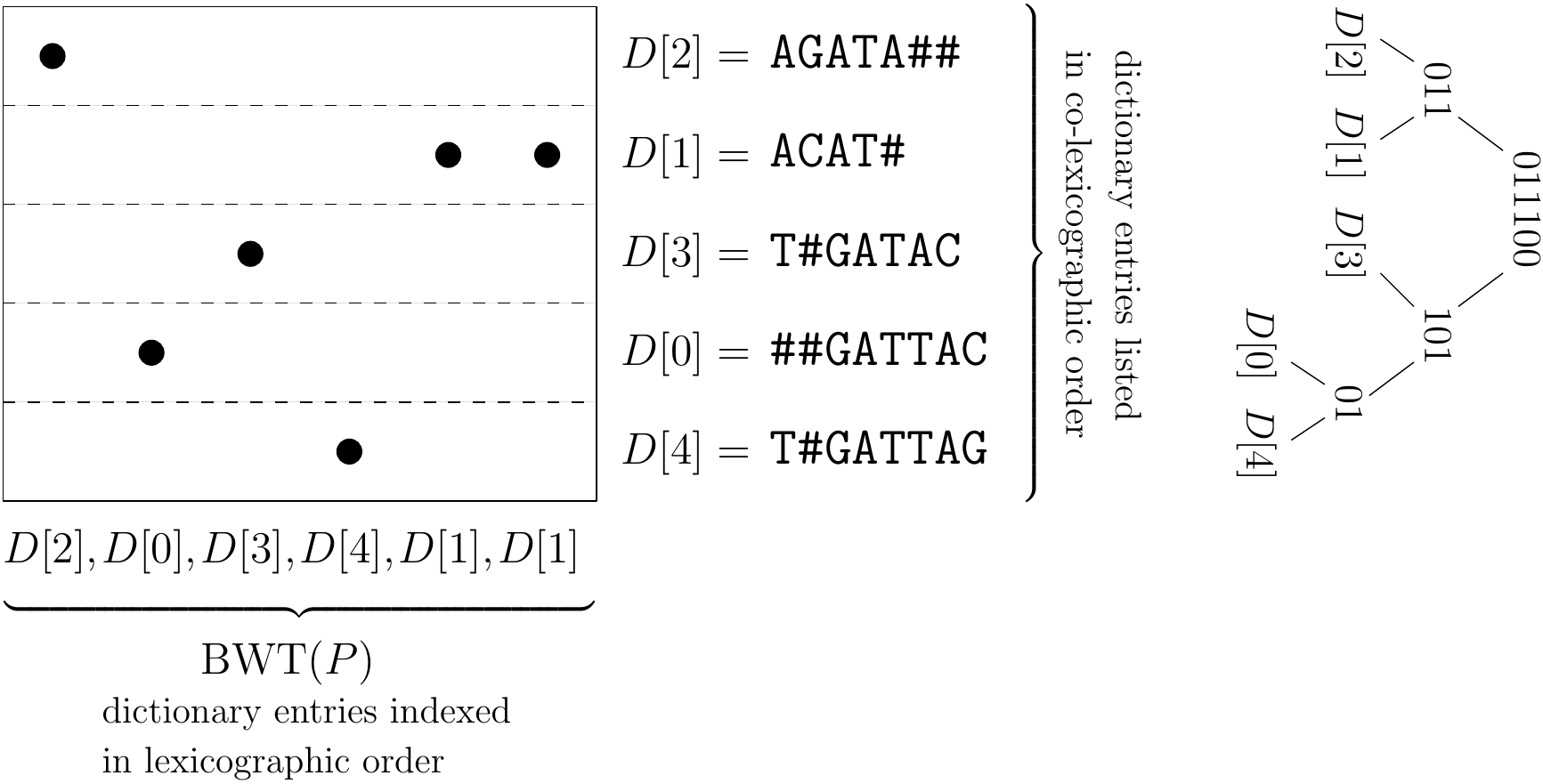}
\caption{The wavelet tree $W$ for our example {\bf (right)} and the grid it represents {\bf (left)}.  We have rotated $W$ so its leaves line up with the corresponding rows of the grid.}
\label{fig:grid}
\end{figure}

Fifth, we store a wavelet tree $W$ over the BWT of the phrase identifiers in $P$, with the leaves of the wavelet tree labelled from left to right by the set of phrase identifiers in co-lexicographic order.  Labelling the leaves of $W$ this way means that leaves labelled with identifiers of phrases ending with the same suffix $\alpha$, are consecutive.  Figure~\ref{fig:grid} shows $W$ for our example, as a wavelet tree and as a grid.

We can use $W$ for 2-sided range counting queries (see~\cite{DBLP:journals/jda/Navarro14}): given $j$ and $r$, we can say how many of the first $j$ phrases in the BWT of $P$ have co-lexicographic rank at least $r$.  We can implement a 3-sided range counting query with two 2-sided range counting queries, and we can implement a 3-sided range selection query using binary search with a 3-sided range counting query at each step of the search.  That means that, given a co-lexicographic range and a value $j$, we can return the index of the $j$th phrase in that interval to appear in the BWT of $P$.

Finally, we store the permutation $\pi$ that maps phrases from their positions in the BWT of $P$ to their positions in $P$ itself.  In our example, $\pi (0) = 5$, $\pi (1) = 0$, $\pi (2) = 2$, $\pi (3) = 4$, $\pi (4) = 1$ and $\pi (5) = 3$.

All these structures take a total of $O (|\PFP (S)|)$ space when we represent $B_P$ and $B_{\BWT}$ with sparse bitvectors --- we note that the number of 1s in $B_{\BWT}$ is bounded by the number of phrase suffixes in the dictionary --- and can be built relatively easy from $P$ and $D$.

\section{Queries}
\label{sec:queries}

\subsection{Access queries}
\label{subsec:access}

The simplest query we consider is random access to $S$.  To find $S [i]$ when given $i$, we use $B_P$ to find the index of the phrase containing $S [i]$ and $S [i]$'s offset in that phrase.  We then use random access to $P$ to identify that phrase, and random access to $D$ to return the appropriate character.

\subsection{LCE queries}
\label{subsec:LCE}

A longest common extension (LCE) query $\LCE (i, j)$ should return the length of the longest common prefix of $S [i..n - 1]$ and $S [j..n - 1]$.  In our example, $\LCE (3, 11) = 9$ because the longest common prefix of $\mathtt{TACAT\#GATACAT\#GATTAGATA\#\#}$ and $\mathtt{TACAT\#GATTAGATA\#\#}$ is $\mathtt{TACAT\#GAT}$.

Given $i$ and $j$, we use the bitvector $B_P$ to find the phrases containing $S [i]$ and $S [j]$ and their offsets in those phrases.  Let $\alpha$ and $\beta$ be the suffixes of those phrases starting at $S [i]$ and $S [j]$, so $|\alpha|, |\beta| > w$.  In our example, the phrases containing $S [3]$ and $S [11]$ are $S [0..5] = \mathtt{GATTAC}$ and $S [9..13]  = \mathtt{GATAC}$.

By Lemma~\ref{lem:prefix-freeness}, neither $\alpha$ nor $\beta$ is a proper prefix of the other, so there are only the following two possibilities: first, $\alpha [k] \neq \beta [k]$ for some $k < |\alpha|, |\beta|$, so $\LCE (i, j)$ is the length of the longest common prefix of $\alpha$ and $\beta$; second, $\alpha = \beta$, so
\[\LCE (i, j) = |\alpha| + \LCE (i + |\alpha|, j + |\alpha|)\,\]
where $S [i + |\alpha|..n - 1]$ and $S [j + |\alpha|..n - 1]$ are both suffixes of $S$ starting immediately after trigger strings.  In our example, $\alpha = \beta = \mathtt{TAC}$, so $\LCE (3, 11) = 3 + \LCE (6, 14)$.

There are several ways we can find the length of the longest common prefix of phrase suffixes quickly using $O (|\PFP (S)|)$ space, such as storing an inverse suffix array (ISA) of the dictionary $D$ of distinct phrases, an LCP array for the dictionary, and a data structure supporting range-minimum queries (RMQs) over that LCP array.  This takes $O (|\PFP (S)|)$ space.  Perhaps the most practical option, however, is to simply compare the phrase suffixes machine-word by machine-word until finding a mismatch.

To find the length of the longest common prefix of two suffixes of $S$ starting immediately after trigger strings, we can simply store a hash table mapping the starting position of each such suffix to its lexicographic rank among such suffixes, an LCP array for those suffixes, and an RMQ data structure over that LCP array.  This also takes $O (|\PFP (S)|)$ space.

For our example, the hash table maps 0 to 4, 6 to 0, 9 to 3, 14 to 1, 17 to 5, and 23 to 2.  The suffixes of $S$ starting after trigger strings and the their LCP values are shown below:
\begin{eqnarray*}
0 && \mathtt{AT\#GATACAT\#GATTAGATA\#\#}\\
6 && \mathtt{AT\#GATTAGATA\#\#}\\
2 && \mathtt{ATA\#\#}\\
0 && \mathtt{GATACAT\#GATTAGATA\#\#}\\
3 && \mathtt{GATTACAT\#GATACAT\#GATTAGATA\#\#}\\
5 && \mathtt{GATTAGATA\#\#}
\end{eqnarray*}
In our example query, having reduced computing $\LCE (3, 11)$ to computing $\LCE (6, 14)$, we map 6 to 0 and 14 to 1 and use an RMQ to find the length 6 of the longest common prefix of $\mathtt{AT\#GATACAT\#GATTAGATA\#\#}$ and $\mathtt{AT\#GATTAGATA\#\#}$.

\subsection{SA queries}
\label{subsec:SA}

A suffix array (SA) query $\SA [i]$ should return the starting position in $S$ (counting from 0) of its lexicographically $i$th suffix or, equivalently, the number 1 greater than the position in $S$ of $\BWT [i]$.  In our example, $\SA [24] = 11$ because the suffix of $S$ with lexicographic rank 24 (counting from 0) is $S [11..27] = \mathtt{TACAT\#GATTAGATA\#\#}$.

Given $i$, we use $B_{\BWT}.\rank (i) - 1$ and $i - B_{\BWT}.\select (B_{\BWT}.\rank (i))$ to find the lexicographic rank (counting from 0) of the proper phrase suffix $\alpha$ of length at least $w$ that starts at $\SA [i]$, and the lexicographic rank $j$ (counting from 0) of $S [\SA [i]..n - 1]$ among the suffixes of $S$ starting with $\alpha$.  In our example, $B_{\BWT}.\rank (24) - 1 = 19$ and $j = 24 - B_{\BWT}.\select (20) = 1$.

We check $M$ to find the length of $\alpha$ and the lexicographic range of the reversed phrases starting with $\alpha$ reversed or, equivalently, the co-lexicographic range of the phrases ending with $\alpha$.  We use $W$ to find the index $k$ of the $j$th phrase in that co-lexicographic interval to appear in the BWT of $P$.  In our example, $M [19] = (3, [2, 3])$ and so $k = 2$.

Since $\alpha$ has length at least $w$, all of its occurrences in $S$ are phrase suffixes.  By Lemma~\ref{lem:comparison}, the lexicographic order of the suffixes of $S$ starting with $\alpha$, is the same as the lexicographic order of the parses starting at the trigger strings that are the last $w$ characters of each of those occurrences of $\alpha$.

Since the lexicographic order of those parses is what determines the order in which the phrases ending with $\alpha$ appear in the BWT of $P$, mapping the $k$th phrase of the BWT of $P$ to its position in $P$ tells us which phrase in $P$ contains the starting point of the lexicographically $j$th suffix of $S$ starting with $\alpha$.  Since we know the length of $\alpha$ from $M$, we can use $B_P$ to find $\SA [i]$.

In our example, since $\pi (2) = 2$ and the first component of $M [19]$ is 3, we know $S [\SA [24]$ is the third to last character in $P [2]$.  Since $w = 2$, the corresponding bit of $B_P$ precedes the third 1 (which marks the start of the trigger string at the beginning of $P [3]$), meaning $\SA [i] = 11$.

Inverse suffix array (ISA) queries can be implemented symmetrically: given a position in $S$, we use $B_P$ to determine which phrase it is in and its offset in that phrase, map that phrase from $P$ to the BWT of $P$, find the co-lexicographic range of phrases ending with the suffix following the character in the phrase, use $W$ to determine the phrase's rank in that range, and use $M$ to determine the position in the BWT of $S$ of the preceding character in $S$.  We will give more details in the full version of this paper.

\subsection{LCP and BWT queries}
\label{subsec:LCPandBWT}

The longest common prefix (LCP) array $\LCP [0..n - 1]$ for $S$ is defined such that $\LCP [0] = 0$ and $\LCP [i] = \LCE (\SA [i - 1], \SA [i])$ for $1 \leq i < n$, so we can implement LCP queries with LCE and SA queries.  In fact, since $\min_{\ell \leq i \leq r} \{\LCP [i]\} = \LCE (\SA [\ell - 1], \SA [r])$, we can even implement range-minimum queries (RMQs) over the LCP array this way.

Obviously we can return $\BWT [i]$ by computing $\SA [i]$ and then returning $S [\SA [i] - 1]$, but in fact we need not apply $\pi$ since, once we find the index of the phrase containing $\BWT [i]$ in the BWT of $P$, we can extract $\BWT [i]$ directly from $D$.

\section{Experiments}
\label{sec:experiments}

We implemented the data structures and measured their performance on real-world datasets. Experiments were performed on a server with Intel(R) Xeon(R) CPU E5-2640 v4 @ 2.40GHz with $40$ cores and 756 gigabytes of RAM running Ubuntu 16.04 (64bit, kernel 4.4.0). The compiler was {\tt g++} version 5.4.0 with {\tt -O3} {\tt -DNDEBUG} {\tt -funroll-loops} {\tt -msse4.2} options. Runtimes were recorded with the C++11 {\tt high\_resolution\_clock} facility. The source code is available online at: \url{https://github.com/maxrossi91/pfp-data-structures}.

\subparagraph{Data}
We used real-world datasets from the Pizza\&Chili repetitive corpus
~\cite{pizzachili}, {\it Salmonella} genomes taken from the GenomeTrakr project~\cite{stevens2017public}, and human chromosome 19 genomes from the 1000 Genomes Project~\cite{1000genomes}; see Table~\ref{tab:realdatasets}.  The Pizza\&Chili repetitive corpus is a collection of repetitive texts characterized by different lengths and alphabet sizes.  GenomeTrakr is an international project dedicated to isolating and sequencing foodbourne pathogens, including Salmonella. Hence, we used $6$ collections of $50$, $100$, $500$, $1000$, $5000$, and $10000$ Salmonella genomes taken from GenomeTrakr. Lastly, we used $10$ sets of variants of human chromosome 19 ({\tt chr19}), containing $1$, $2$, $4$, $8$, $16$, $32$, $64$, $128$, $256$, $512$, and $1000$ distinct variants respectively. Each collection is
a superset of the previous.

	 	         \begin{table}%
	 		\centering
 			\scalebox{.8}{%
    	 		\sisetup{detect-weight = true,
    	 		detect-inline-weight = text,
    	 		table-number-alignment = right,
    	 		round-mode=places,
    	 		round-precision=2}
		 		\begin{tabular}{llS[table-format=3.0]S[table-format=5.2]S[table-format=4.2]S[table-format=4.2]S[table-format=4.2]}
		 			\hline
		 			Name & Description & $\sigma$ & {$n/10^6$} & {$n/r$} & {Dict. (MB)} & {Parse (MB)}\\
		 			\hline
		 			{\tt cere} & Baking yeast genomes & 5 & 461.286644 & 157.19 & 75.611569   & 62.113700 \\
		 			{\tt einstein.de.txt} &  Wikipedia articles in German & 117 & 92.213503 & 5216.14 & 0.684906 & 16.126432 \\
		 			{\tt einstein.en.txt} & Wikipedia articles in English & 139 & 465.248293 & 8961.42 & 1.663777 & 61.440344\\
		 			{\tt Escherichia\_Coli} & Bacteria genomes & 15 & 112.689515 & 32.83 & 42.785576  & 17.441804 \\
		 			{\tt influenza} & Virus genomes  & 15 & 154.808555 & 251.3 & 37.553971  & 2.5184936  \\
		 			{\tt kernel} & Linux Kernel sources  & 160 & 249.514406 & 499.82 & 11.132045  & 45.937840  \\
		 			{\tt para} & Yeast genomes  & 5 & 429.265758 & 111.78 & 72.985087  & 75.079952  \\
		 			{\tt world\_leaders} & CIA world leaders files  & 89 & 46.905202 & 634.9 & 7.028739  & 3.496608 \\
		 			{\tt chr19.1000} & Human chromosome 19  & 5 & 60110.548090 & 1287.38 & 274.626513  & 2219.075272 \\
		 			{\tt Salmonella.10000} & Salmonella genomes database  & 4 & 51820.383427 & 36.61 & 4483.430006  & 2039.160816 \\
		 			\hline
		 		\end{tabular}
		 	}
	 		\caption{Datasets used in the experiments.  We give the names and descriptions of the datasets in the first two columns. In column 3 we give the alphabet size. In columns 4 and 5 we report the length of the file and the ratio of the length to the number of runs in the BWT. Lastly, we give the size of the dictionary and the parse in columns 6 and 7, respectively.\label{tab:realdatasets}}
	 	\end{table}

\subparagraph{Data structures}

We compared the PFP data structures implementation ({\tt pfp.ds}); the compressed suffix tree implementation ({\tt sdsl.cst}) from the {\tt sdsl-lite} library~\cite{gbmp2014sea}; and the block tree compressed suffix tree implementation ({\tt bt.cst}) of C{\'{a}}ceres and Navarro \cite{DBLP:conf/spire/CaceresN19}.

\subparagraph{Implementation}

We implemented the PFP data structures using {\tt sdsl-lite} library~\cite{gbmp2014sea} bitvectors and their rank and select supports. We used SACA-K~\cite{DBLP:journals/tois/Nong13} to sort the parse lexicographically, and gSACA-K~\cite{LouzaGT17b} to compute the SA, LCP array and document array of the dictionary.

Using gSACA-K to sort the dictionary, we can use the same phrase terminator to concatenate each phrase. The result of gSACA-K is equivalent to the result obtained if we concatenate unique terminators in a lexicographically increase order, as required for the computation of $B_{BWT}$.

\subparagraph{Building test setup}
We tested the running time and peak memory usage of the data structures during the building. For building the PFP data structures, we first computed the prefix free parsing of the dataset using BigBWT~\cite{DBLP:journals/almob/BoucherGKLMM19} with 32 threads, a window size $w=10$, and parameter $p=100$. The resulting output is loaded in memory and used to build the PFP data structures. The running time for the construction of the PFP data structures includes the time to build the parse as well as the time to store the parse to disk.

We built each data structure $5$ times for the Pizza\&Chili corpus datasets, for the sets of chromosome $19$ up to $64$ distinct variants, and  for Salmonella up to $1000$ sequences. The remaining experiments have been tested only once. The experiments that exceeded 15 hours were omitted from further consideration, e.g. {\tt chr19.1000} and {\tt salmonella.10000} for {\tt sdsl.cst}.  Furthermore, {\tt bt.cst} failed to successfully build for the sets of chromosome $19$ greater than $16$ distinct variants, and for Salmonella with more than $100$ sequences due to integer overflows causing segmentation fault errors.

\subparagraph{Querying test setup}
We performed suffix array (SA), longest common prefix (LCP), and longest common extension (LCE) queries on each data structure. We performed the queries only for chromosome $19$ and salmonella datasets. We used google benchmarks\footnote{\url{https://github.com/google/benchmark}} for query testing. 

For the SA and LCP queries, we generated $10000$ randomly distributed indices in the data structures, and queried the data structure. Similarly, for the LCE queries, we generated $10000$ randomly distributed pairs of indices in the data structures, and queried the data structure. The running time for the queries is reported in average over the $10000$ access.  Lastly, we note that we aborted any query that exceeded 15 hours, and therefore, omitted  the LCE queries for {\tt bt.cst}.

\subparagraph{Time}

\begin{figure}[t]
    \centering
 	\begin{subfigure}[b]{.45\textwidth}
 		\includegraphics[width=\textwidth]{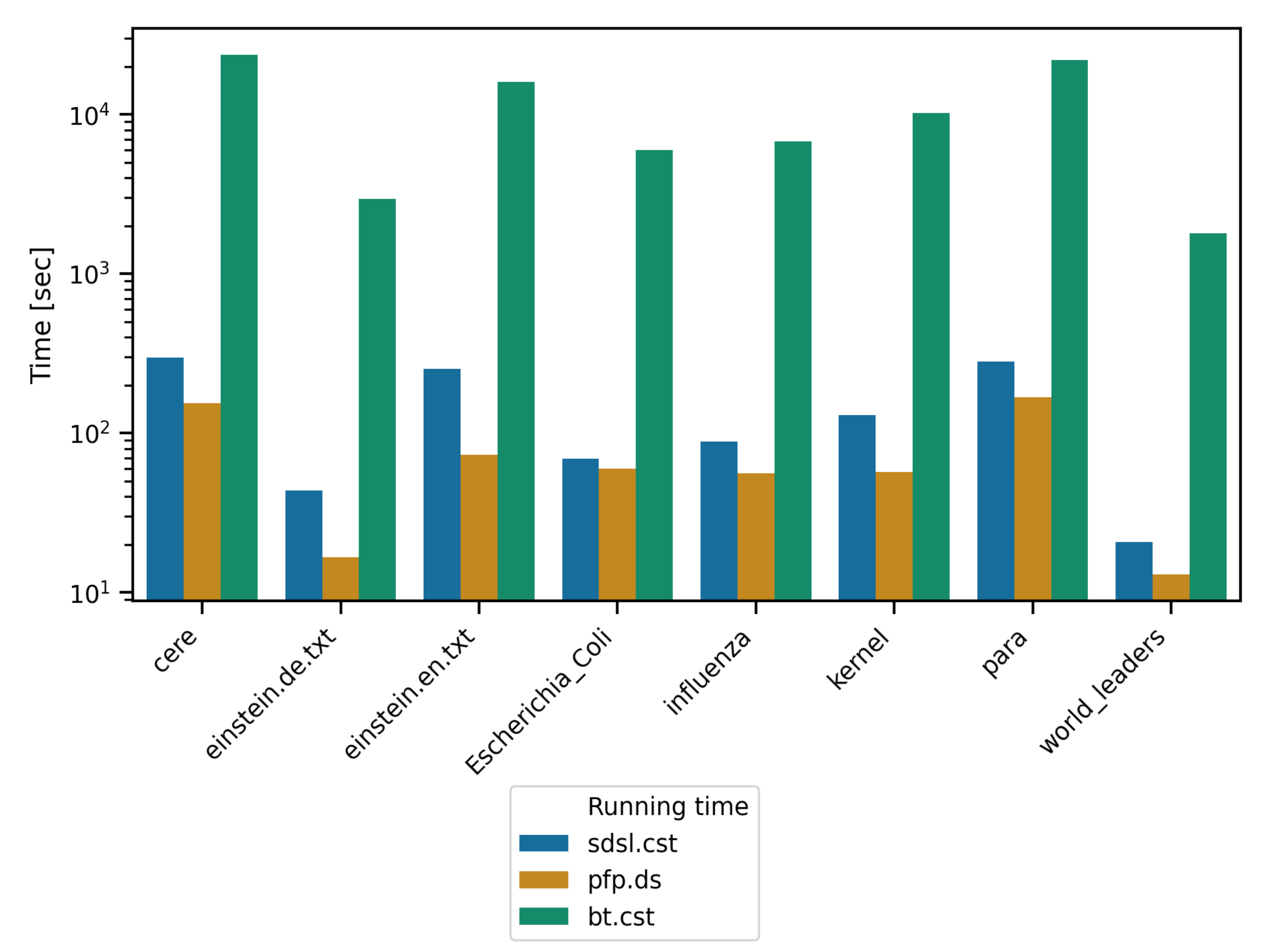}
 		\caption{Construction time\label{fig:repcorpus:time}}
 	\end{subfigure} 	
 	\begin{subfigure}[b]{.45\textwidth}
 		\includegraphics[width=\textwidth]{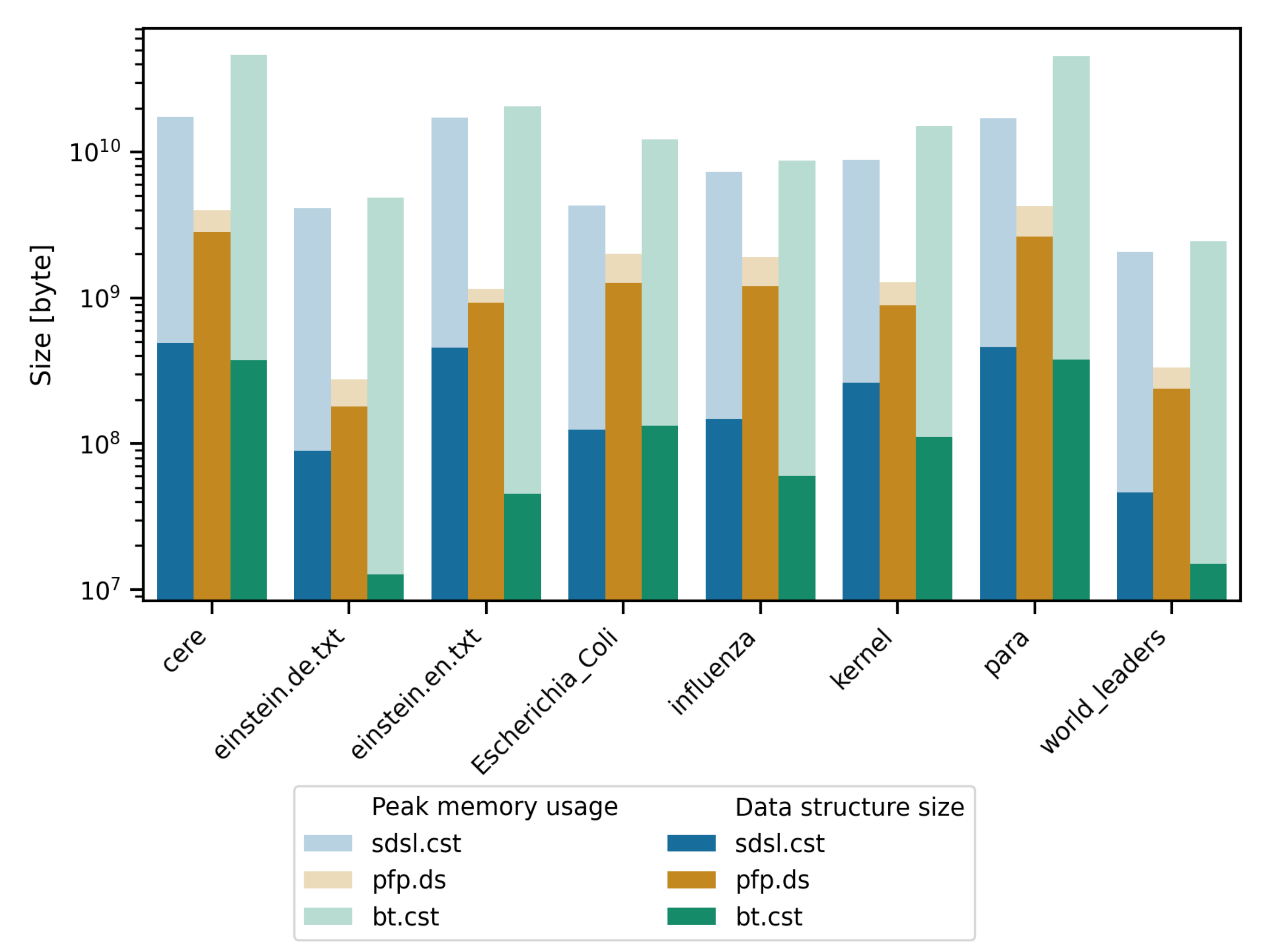}
 		\caption{Peak memory usage and data structure size\label{fig:repcorpus:space}}
 	\end{subfigure}
  	\caption{Pizza\&Chili dataset construction running time (\ref{fig:repcorpus:time}), peak memory usage and data structure size (\ref{fig:repcorpus:space}). We compare the prefix-free parsing data structure ({\tt pfp.ds}) with the sdsl compressed suffix tree ({\tt sdsl.cst}), and with the block tree based compressed suffix tree ({\tt bt.cst}). \label{fig:repcorpus}}
\end{figure}

\begin{figure}[t]
    \centering
 	\begin{subfigure}[b]{.49\textwidth}
 		\includegraphics[width=\textwidth]{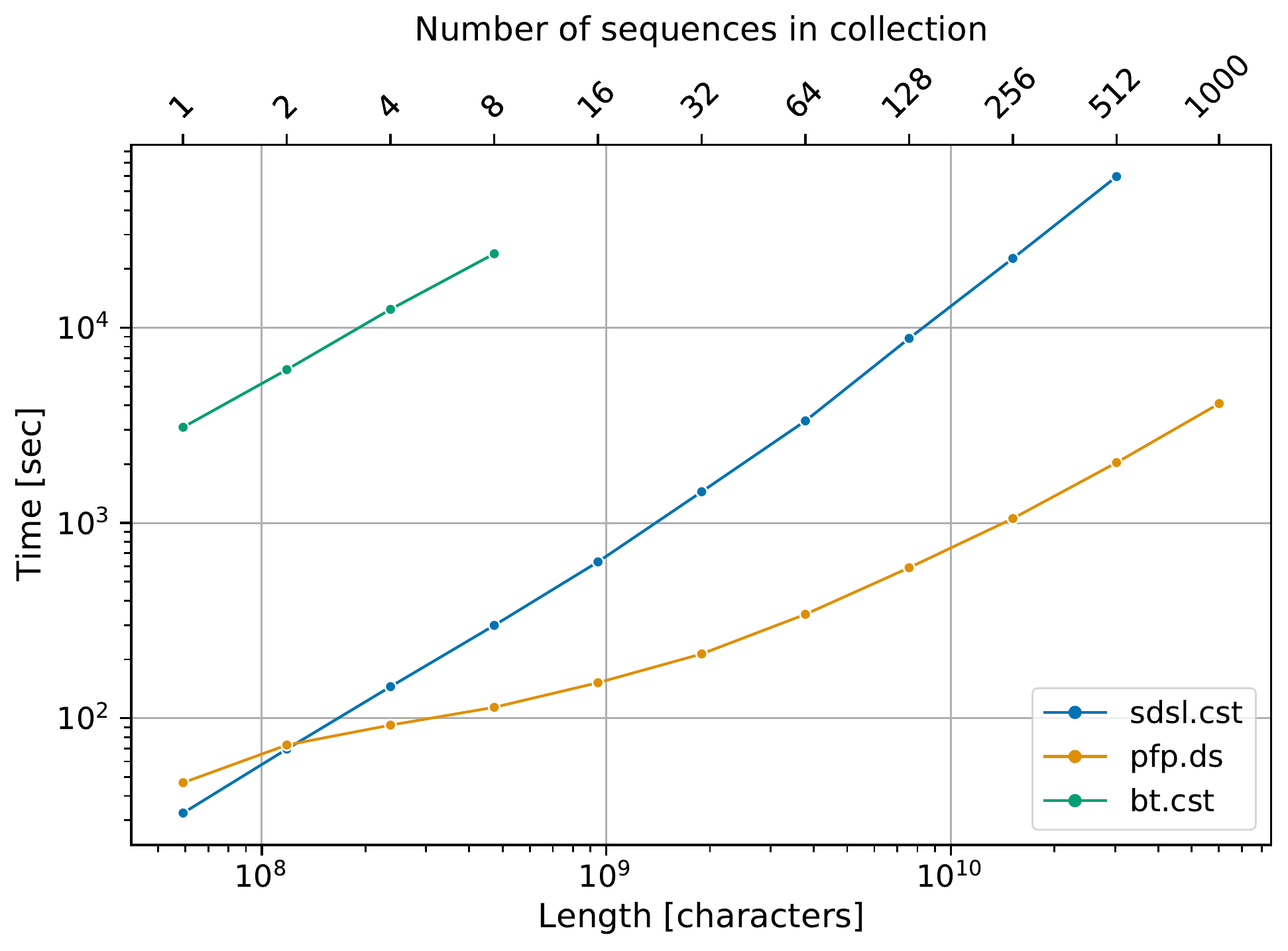}
 		\caption{Construction time\label{fig:chr19:time}}
 	\end{subfigure} 	
 	\begin{subfigure}[b]{.49\textwidth}
 		\includegraphics[width=\textwidth]{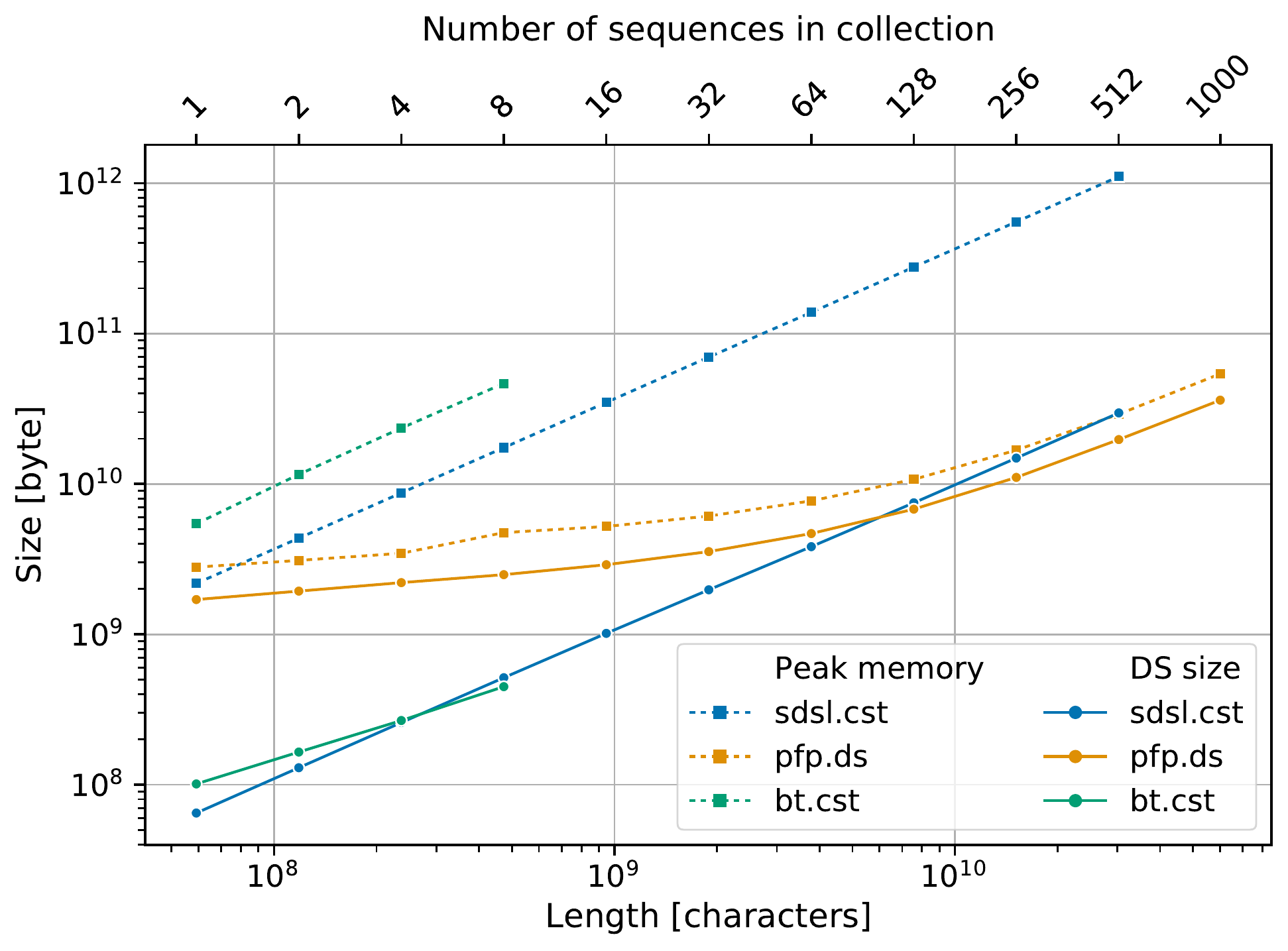}
 		\caption{Peak memory and data structure size\label{fig:chr19:space}}
 	\end{subfigure}
 	\begin{subfigure}[b]{.49\textwidth}
 		\includegraphics[width=\textwidth]{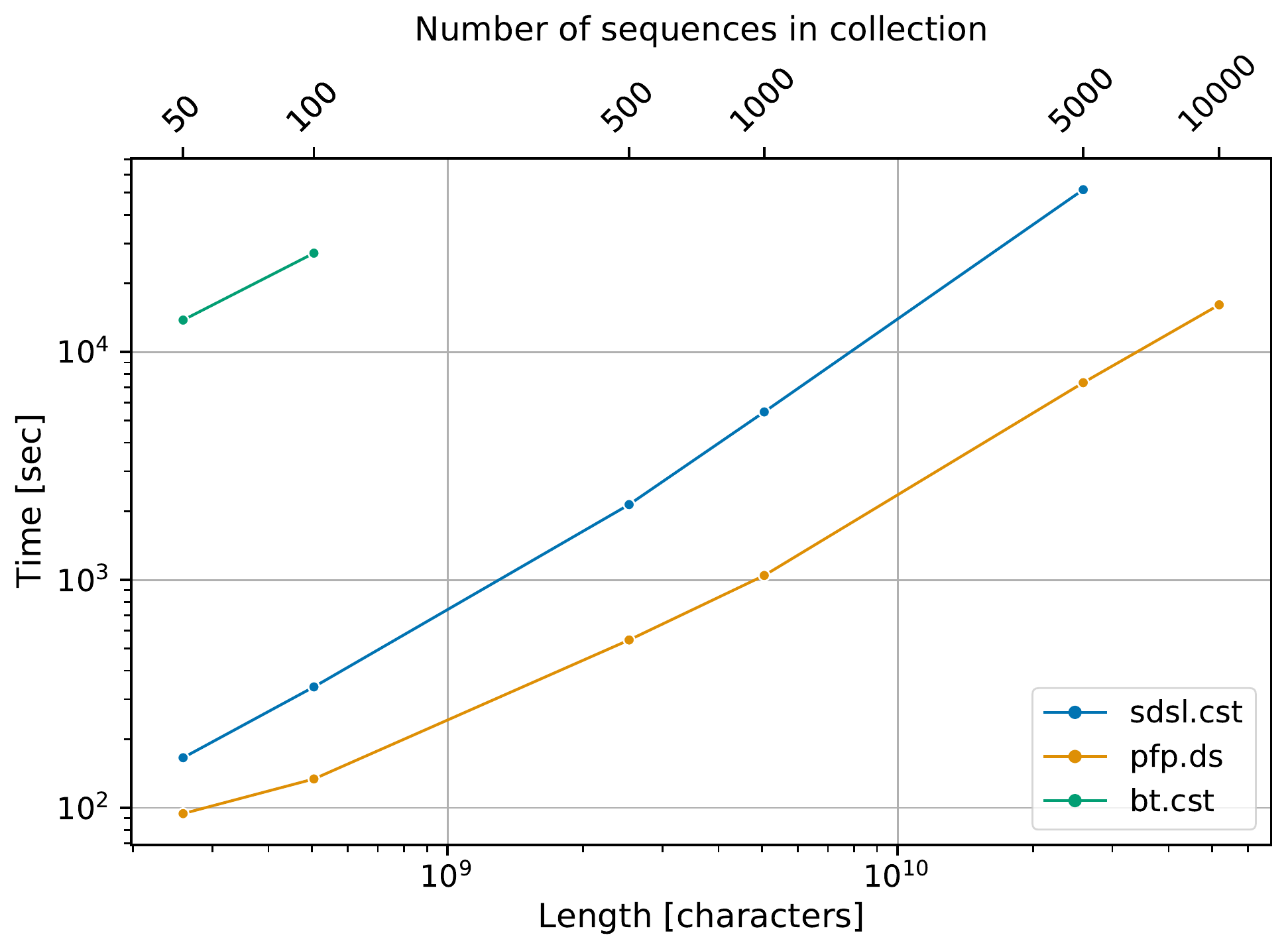}
 		\caption{Construction time\label{fig:salmonella:time}}
 	\end{subfigure} 	
 	\begin{subfigure}[b]{.49\textwidth}
 		\includegraphics[width=\textwidth]{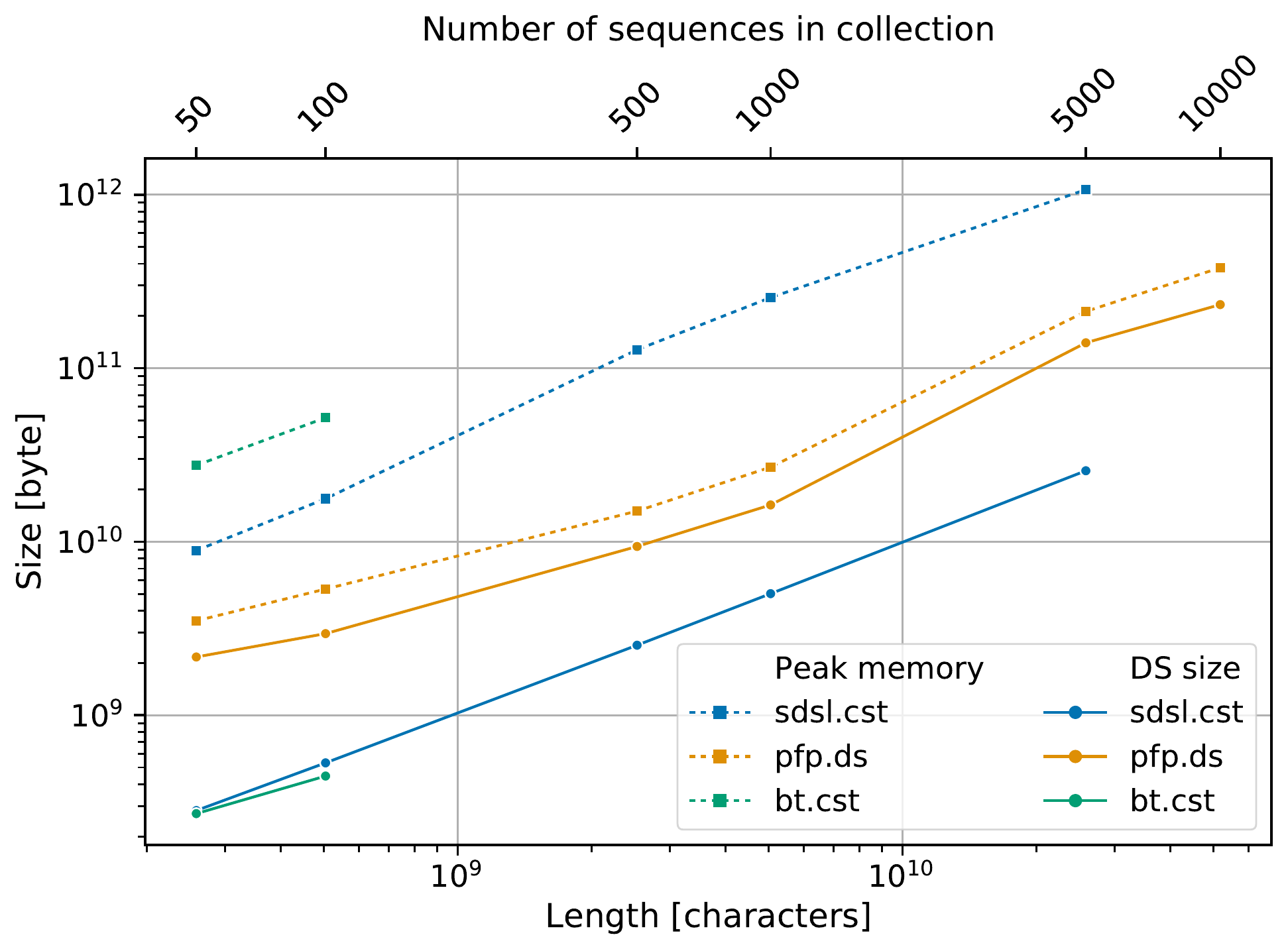}
 		\caption{Peak memory and data structure size\label{fig:salmonella:space}}
 	\end{subfigure}
 	\caption{Chromosome 19, and salmonella datasets construction time (~\ref{fig:chr19:time},~\ref{fig:salmonella:time}), peak memory usage and data structure size (~\ref{fig:chr19:space},~\ref{fig:salmonella:space}). We compare the prefix-free parsing data structure ({\tt pfp.ds}) with the sdsl compressed suffix tree ({\tt sdsl.cst}), and with the block tree based compressed suffix tree ({\tt bt.cst}). \label{fig:chr19}\label{fig:salmonella}}
\end{figure}

Figures~\ref{fig:repcorpus:time}, \ref{fig:chr19:time}, and ~\ref{fig:salmonella:time} illustrate the construction time for all the data structures for Pizza\&Chili, Chromosome 19, and Salmonella, respectively.  From the reported data, we observe that the construction of {\tt pfp.ds} is always faster than {\tt sdsl.cst} and {\tt bt.cst}, except for the cases {\tt chr19.1} and {\tt chr19.2} in which {\tt sdsl.cst} is the fastest to be built.  The maximum speedup of {\tt pfp.ds} with respect to {\tt sdsl.cst} is 3.47x ({\tt einstein.en.txt}), 29.19x ({\tt chr19.512}), and 7.68x ({\tt salmonella.5000}). The speedup of {\tt pfp.ds} with respect to {\tt bt.cst} is 219x ({\tt einstein.en.txt}), 210x ({\tt chr19.8}), and 202x ({\tt salmonella.100}).

From Figure~\ref{fig:chr19:time} we observe that doubling the length of the dataset, the running time of {\tt pfp.ds} increases by a factor of $1.9$ when moving from $256$ variants to $512$, and a factor of $2$ when moving from $512$ variants to $1000$. On the other hand {\tt sdsl.cst} running time increases by a factor of $2.6$ when moving from $256$ variants to $512$.

From Figure~\ref{fig:salmonella:time} we observe that increasing the length of the dataset by a factor of $10$, when moving from $500$ to $5000$, it increases the running time of  {\tt pfp.ds} by a factor of $12$, while for {\tt sdsl.cst} it increases by a factor of $24$. 

\subparagraph{Space}
Figures~\ref{fig:repcorpus:space}, \ref{fig:chr19:space}, and ~\ref{fig:salmonella:space} illustrate the memory peak usage and the size of the data structure for all the data structures for Pizza\&Chili, Chromosome 19, and Salmonella, respectively. We observe that the peak memory usage of {\tt pfp.ds} is almost always less than both {\tt sdsl.cst} and {\tt bt.cst}. Yet, the size of {\tt pfp.ds} is larger than  both {\tt sdsl.cst} and {\tt bt.cst}, except for {\tt chr19.128}, {\tt chr19.256}, and {\tt chr19.512}, where {\tt pfp.ds} is the smallest one.

\begin{figure}[t]
    \centering
    \includegraphics[width=.99\textwidth]{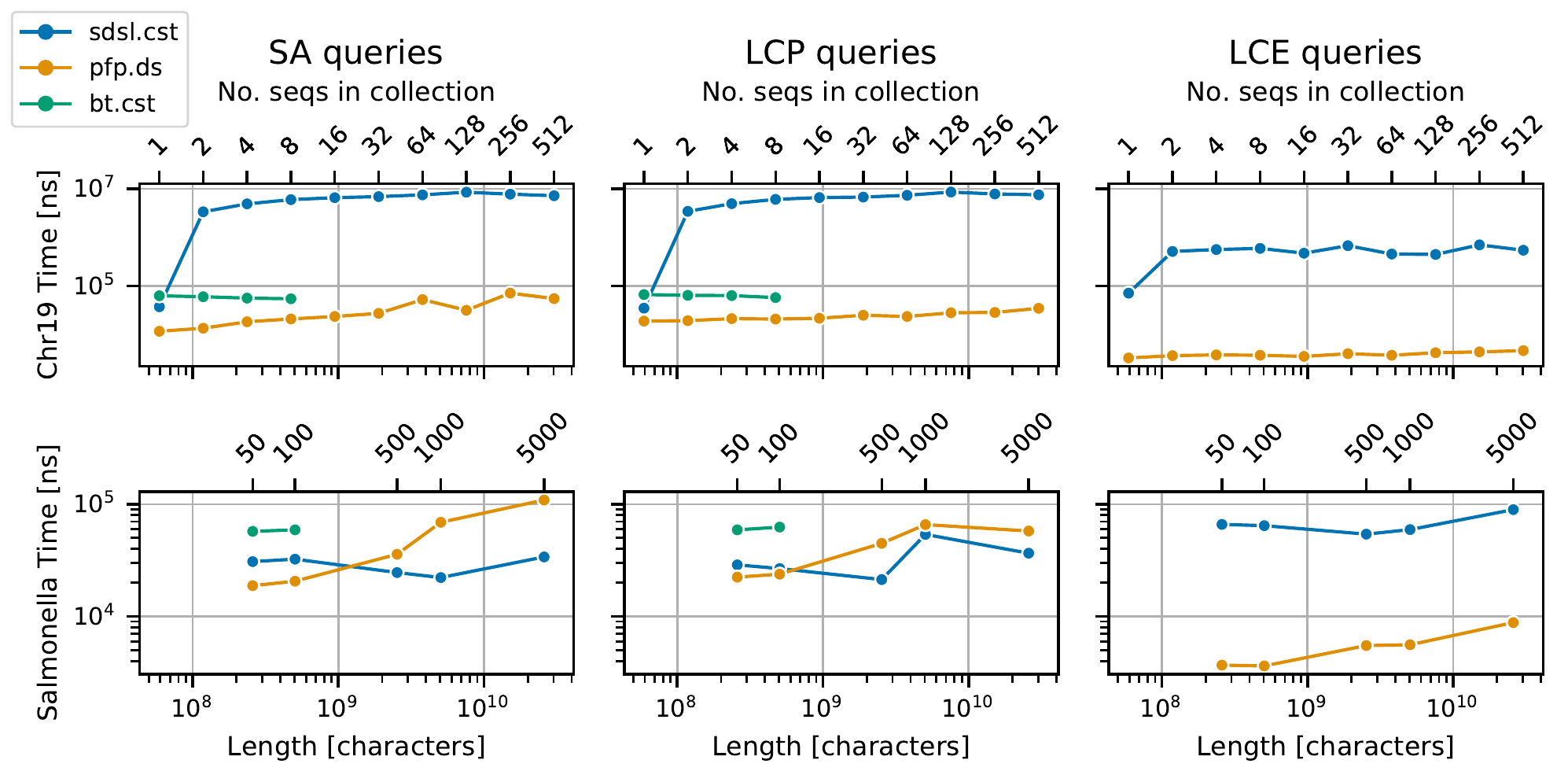}
    \caption{Running time, Peak memory usage and data structure size for suffix array, longest common prefix and longest common extension queries for the chromosome 19 datasets and the salmonella datasets. \label{fig:queries}}
\end{figure}

We can also observe that the difference between the memory peak usage and the data structure size is very small in {\tt pfp.ds}. Its maximum ratio is attained at {\tt chr.16} where the memory peak is 1.9x larger than the data structure size. For the Pizza\&Chili dataset the maximum is 1.6x for {\tt para}, while for the salmonella dataset the maximum is 1.8x for {\tt salmonella.100}.

Moreover, the change of trend in the memory usage of {\tt pfp.ds} from {\tt salmonella.1000} to {\tt salmonella.5000} is because we switched from gSACA-K $32$ bit version to gSAKA-K $64$ bit version, since the 32 bit version can sort text of length up to 2GB and the length of the dictionary is larger than 2GB.

\subparagraph{Queries}
In Figure~\ref{fig:queries} we report the time that each data structure to perform SA, LCP, and LCE queries. We observe that {\tt pfp.ds} is always faster than {\tt sdsl.cst} on LCE queries. On chromosome 19 dataset, {\tt pfp.ds} is faster than {\tt sdsl.cst} and {\tt bt.cst} on SA and LCP queries.

On the other hand, on Salmonella dataset, {\tt sdsl.cst} is faster than {\tt pfp.ds} on SA queries and LCP queries for more than 100 genomes.

\end{document}